\documentclass[aps,prl,floatfix,epsfig,twocolumn,showpacs,preprintnumbers]{revtex4}
\usepackage{amssymb}
\usepackage{graphicx}
\usepackage{amsmath}

\input{tcilatex}
\begin{document}

\title{Computational Design of Axion Insulators Based on 5\textit{d} Spinels
Compounds}
\author{Xiangang Wan$^{1}$, Ashvin Vishwanath$^{2,3}$ and Sergey Y. Savrasov$%
^{4}$}
\affiliation{$^{1}$National Laboratory of Solid State Microstructures and Department of
Physics, Nanjing University, Nanjing 210093, China}
\affiliation{$^{2}$Department of Physics, University of California, Berkeley, CA 94720}
\affiliation{$^{3}$ Materials Sciences Division, Lawrence Berkeley National Laboratory,
Berkeley CA 94720.}
\affiliation{$^{4}$Department of Physics, University of California, Davis, One Shields
Avenue, Davis, CA 95616.}

\begin{abstract}
Based on density functional calculation with LDA+U\ method, we propose that
hypothetical Osmium compounds such as CaOs$_{2}$O$_{4}$ and SrOs$_{2}$O$_{4}$
can be stabilized in the geometrically frustrated spinel crystal structure.
They also show some exotic electronic and magnetic properties in a
reasonable range of on--site Coulomb correlation $U$ such as ferromagnetism
and orbital magnetoelectric effect characteristic to Axion electrodynamics.
Other electronic phases including 3D Dirac metal and Mott insulator exist
and would make perspective 5d spinels ideal for applications.
\end{abstract}

\date{\today}
\maketitle

\input{epsf}

Recently, a theoretical work has played a key role in discovery of novel
topological insulators (TI), where time reversal symmetry protects unusual
surface states which could lead to electronic devices with new
functionalities \cite{Qi-Zhang Phys.Today,Hasan-review,Moore-2010}.\ Unlike
other fields of condensed matter physics, where most compounds are found
haphazardly, here a number of new materials has been suggested based on band
structure calculations \cite{Zhang-2006,FuKane,Fang-Dai-Zhang}, some of
which have been confirmed experimentally \cite{BiSb-exp,Bi2Se3-exp}.

In addition to these unusual electronic properties, introducing magnetic
order on the surface of TIs can lead to a gap for surface states, leading to
completely insulating behavior. Here, novel phenomena such as the half
quantum Hall effect \cite{FuKaneMele} and quantized magnetoelectric effect 
\cite{Zhang} appear. The latter is parameterized by the quantized value of
the Axion electrodynamics parameter $\theta =\pi $, \cite{Zhang, Wilczek}.
When time reversal symmetry, as well as spatial symmetries like inversion
are broken in the bulk, the $\theta $ parameter is no longer quantized\cite%
{Axion-3,AndrewIvo}. However, in most known materials, it is extremely small
eg. $\theta =10^{-3},\,10^{-4}$ in Cr$_{2}$O$_{3}$ and BiFeO$_{3}$
respectively \cite{AndrewIvo}. Thus, achieving large values of the $\theta $
parameter, even in the absence of quantization, is an important materials
challenge. For sufficiently large $\theta $, $\theta >\pi /2$, it has been
proposed that magnetic domain walls will be associated with conducting
channels at the sample surface\cite{TeoKane}. Recent experimental progress
in introducing magnetism via dopants \cite{Hasan,ZXShen} in Bi$_{2}$Se$_{3}$
has been reported, as well as tuning of chemical potential into the small
gap induced on surface states \cite{ZXShen}. Ideally however, magnetic
topological insulators would combine band topology with intrinsic magnetic
order, leading to large surface energy gaps. Unfortunately, most of
known/proposed TIs\ involve \textit{p}-electron orbitals, whose Coulomb
interaction is weak and cannot support magnetism. Elements with 3\textit{d}
and 4\textit{d} electrons do have large electronic correlations, but only
small SOC. While both strong Coulomb interaction and large SOC is found in 4%
\textit{f} and 5\textit{f} electron systems, they usually form narrow energy
bands, thus making it also hard to realize TI state of matter (see \cite%
{Coleman} for an exception).

Due to the interplay of strong SOC and moderate electronic correlations,
very recently systems with 5\textit{d} electrons have received a lot of
research attention\cite{Iridates,Sr2IrO4-1,Spin-liquid
Na4I3O8,Pesin-Balents,Kim,Dirac-Metal}. It has been suggested that
pyrochlore iridates exhibit novel phases such, e.g., TI\cite%
{Pesin-Balents,Kim}, and\ topological semi-metal (TSM) \cite{Dirac-Metal}
behavior. As in pyrochlores \textit{A}$_{2}$\textit{B}$_{2}$O$_{7}$, spinel
compounds \textit{AB}$_{2}$O$_{4}$ have their four \textit{B}--sites forming
a corner sharing tetrahedral network. Furthermore, it may be expected that 5%
\textit{d} systems in the spinel structure will be more tunable by pressure,
external fields or by doping as compared to the closely packed pyrochlore
lattice. Previous studies suggested that electronic d$^{5}$ configuration
can realize an insulating state in 5\textit{d} compounds\cite{Sr2IrO4-1}.
Based on valence arguments we therefore focus on spinel osmates \textit{A}Os$%
_{2}$O$_{4}$ (\textit{A} is alkali metal element such as Mg, Ca, Sr, Ba),
and investigate their electronic structure and magnetic properties using
theoretical density functional calculations. Our numerical results reveal
that the spinel structure can at least be a metastable crystallographic
state for \textit{5d} elements which is additionally supported by a very
recent synthesis of an empty\ iridium spinel compound Ir$_{2}$O$_{4}$\cite%
{EmptySpinel}. Most interestingly, we find that the ground state is
ferromagnetic and the electronic properties are sensitive to the \textit{A--}%
site. Depending on Coulomb correlation, they show some exotic properties,
and for the \textit{U} relevant to osmate, we predict CaOs$_{2}$O$_{4}$ and
SrOs$_{2}$O$_{4}$ behave as magnetic axion insulators with $\theta =\pi $.
The quantized value is protected by the inversion symmetry of the lattice.
However, since the surfaces break inversion, typically they are gapped
allowing us to define $\theta $. Unlike other proposals for generating large 
$\theta $, which exploit proximity to a topological insulator state\cite%
{ZXShen}, here a generalized parity criterion\cite{AriZhangAV,Bernevig}
allows us to obtain large $\theta $ even when no proximate TI state exits.

We perform our electronic structure calculations for the spinel Osmates
based on local spin density approximation (LSDA) to density functional
theory (DFT) with the full--potential, all--electron,
linear--muffin--tin--orbital (LMTO) method\cite{FP-LMTO}. Despite the 5%
\textit{d} orbitals are spatially extended, recent theoretical and
experimental work has given the evidence on the importance of Coulomb
interactions in 5\textit{d }compounds\cite{Sr2IrO4-1}. We utilize LSDA+U
scheme\cite{LDA+U} to take into account the effect of Coulomb repulsion, and
vary parameter \textit{U} between 0 and 1.5 eV. We use a 24$\times $24$%
\times $24 k--mesh to perform Brillouin zone integration, and switch off
symmetry operations in order to minimize possible numerical errors in
studies of various (non--)collinear configurations. As the experimental
lattice parameters are not available, we search for the stable crystal
structures by locating the minimum in the calculated total energy as a
function of the lattice constant and internal atomic coordinates.

\begin{figure}[tbp]
\includegraphics [height=2.0in,width=2.5in] {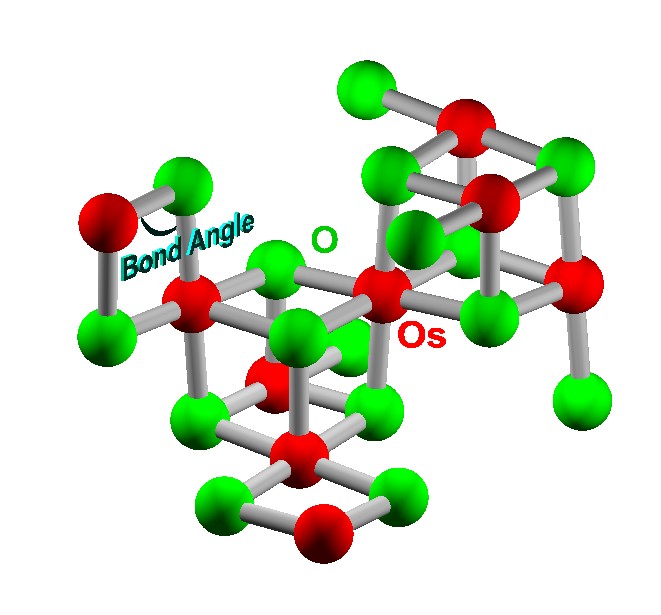}
\caption{Fragment of spinel crystal structure AOs$_{2}$O$_{4}$ (only Os and
O atoms are shown)\ with the Os-O-Os angle being optimized.}
\end{figure}

Spinel structure (see Fig. 1) forms space group \textit{Fd\={3}m}. In order
to allow its relaxation, we change the lattice constant from 13 a.u. to 18
a.u. and vary the Os-O-Os bond angle from 90$^{0}$ to 120$^{0}\ $with a step
of 1$^{0}$. We confirm that the choice of Coulomb \textit{U} has only small
effect on the determination of lattice parameters, and list the lowest
energy structures obtained by LDA+SO (U=0) calculations in Table I. For
comparison, we also list the same parameters for the pyrochlore iridate Y$%
_{2}$Ir$_{2}$O$_{7}$. We see that the \textit{A}-site element has a
considerable effect on the Os--O bond angle and its length. As we discuss
below, this allows us to control the electronic structure in Osmates and
design exotic topological phases.

\begin{table}[ptb]
\caption{Theoretically determined structure parameters of spinel osmates.
Angle denotes the Os-O-Os bond angle; Os-O and Os-Os denote the nearest
neighbor Os-O bond length, and Os-Os bond length, respectively.}%
\begin{tabular}{llll}
\hline
& Angle & Os-Os & Os-O \\ \hline
CaOs$_{2}$O$_{4}$ & 98.5$^{0}$ & 3.07 $\mathring{A}$ & 2.04 $\mathring{A}$
\\ 
SrOs$_{2}$O$_{4}$ & 94.6$^{0}$ & 3.12 $\mathring{A}$ & 2.05$\mathring{A}$ \\ 
BaOs$_{2}$O$_{4}$ & 103.1$^{0}$ & 3.28 $\mathring{A}$ & 2.11$\mathring{A}$
\\ 
Y$_{2}$Ir$_{2}$O$_{4}$ & 129.7$^{0}$ & 3.60 $\mathring{A}$ & 1.99$\mathring {%
A}$ \\ \hline
\end{tabular}%
\end{table}

The 3\textit{s} band of Mg is lower in energy, and appears around the Fermi
level, which makes MgOs$_{2}$O$_{4}$ always metallic. We therefore do not
discuss this compound here and concentrate our study on CaOs$_{2}$O$_{4}.$
Its band structure from non--magnetic LDA+SO calculation is found to be
metallic and shown in Fig.2a. The energy bands around the Fermi level appear
as J$_{eff}$=1/2 states similar to the ones found in Sr$_{2}$IrO$_{4}$\cite%
{Sr2IrO4-1}, and also in Y$_{2}$Ir$_{2}$O$_{7}$\cite{Dirac-Metal}, where a
metal rather than the interesting topological insulator scenario of \cite%
{Pesin-Balents} was obtained due to a 2--4--2 sequence of degeneracies at
the $\Gamma $ point. As importance of electronic correlations for 5\textit{d 
}orbitals has been recently emphasized \cite{Sr2IrO4-1}, we therefore
perform LSDA+U+SO calculations. Although, the accurate value of \textit{U}
is not known, the Os-Os bond length of spinel osmates is shorter than that
of Y$_{2}$Ir$_{2}$O$_{7}$, and one can expect that the \textit{U} in CaOs$%
_{2}$O$_{4}$ is smaller than in Y$_{2}$Ir$_{2}$O$_{7}$. We therefore believe
that the \textit{U} is in the range between 0.5 and 1.5 eV. As in the
pyrochlore structure, the Os spinel sublattice is geometrically frustrated.
Naively one may expect that the magnetic configuration of CaOs$_{2}$O$_{4}$
is also nonconlinear as recently found in Y$_{2}$Ir$_{2}$O$_{7}$\cite%
{Dirac-Metal}. To search for possible magnetic ground states, we perform
calculations by starting with a number of different conlinear and
noncolinear magnetic configurations including ferro-- and antiferromagnetic
(FM/AFM) collinear (010), (110), (111), as well as non--collinear
all-in/all-out, 2-in/2-out, 3-in/1-out and some perpendicular configurations
promoted by Dzyaloshinsky--Moriya interactions\cite{DMPyrochlore}. We find
that when \textit{U} is less than 1.3 eV, only the FM-(010)-configuration
retains its initial input magnetization direction; in all other
configurations the moments depart from their input orientation. We also
consider a two-up, two-down state which is suggested by the strong coupling
limit where Os-O bonds are nearly 90$^{o}$. As shown in \cite{Khalliulin1},
this leads to a 'Kitaev type' ferromagnetic interaction for spin components
perpendicular to the plane. Although this structure is found to be stable
for $U<0.8$eV, it is higher in energy than the ferromagnetic state as shown
in Table II. Regardless of the value of \textit{U}, the FM configuration
with magnetization along (010) is found to be the ground state, and the
energy difference between this and other configurations is quite large.

\begin{table}[ptb]
\caption{The spin $\langle S\rangle$ and orbital $\langle O\rangle$\ moment
(in $\protect\mu_{B}$) as well as the total energy E$_{tot}$ per unit cell
(in meV) for several selected magnetic configurations of CaOs$_{2}$O$_{4}$
as calculated using LSDA+U+SO\ method with U=0.5 eV. ($E_{tot}$ is defined
relative to the ground state.) The IDM is a co-planar configuration
predicted for one sign of Dzyaloshinski Morya interactions in Ref. 
\protect\cite{DMPyrochlore}}
\label{2}%
\begin{tabular}{cccccc}
\hline
Configuration: & (010) & all--in/out & 2--in/2--out & 2--up/2--dn & IDM \\ 
\hline
$\langle$S$\rangle$ & 0.59 & 0.30 & 0.40 & 0.25 & 0.35 \\ 
$\langle$O$\rangle$ & 0.14 & 0.21 & 0.22 & 0.14 & 0.16 \\ 
E$_{tot}$(meV) & 0.00 & 290 & 280 & 518 & 351 \\ \hline
\end{tabular}%
\end{table}

This is understood by examining lattice parameters in Table I, where the
main difference between pyrochlore iridates and spinel osmates is the
Os--O--Os bond angle and the Os--Os bond length. Due to the extended nature
of 5\textit{d} orbitals, the 5\textit{d}--2\textit{p} hybridization is
strong and important for the inter--atomic exchange interaction. In Y$_{2}$Ir%
$_{2}$O$_{7}$, the Ir--O--Ir bond angle is much larger than 90$^{0}$ and
Ir--O--Ir\ antiferromagnetic (AFM) superexchange interaction is dominant.
This results in a strong magnetic frustration and non--conlinear ground
state magnetic configuration\cite{Dirac-Metal}. In contrast to pyrochlores,
the Os--O--Os angle in spinel osmates is close to 90$^{0}$, while Os 5%
\textit{d} orbitals have stronger overlap,\ making a direct ferromagnetic
exchange to be important for CaOs$_{2}$O$_{4}$.

For the values of U=0.5 and 1.5 eV, electronic band structures along high
symmetry lines appear to be insulating as shown in Fig.2b and Fig.2c,
respectively. To check a possibility of the gap closure and the metallic
behavior away from the high symmetry lines of BZ, we perform the calculation
with 100$\times$100$\times$100 k-mesh. This very dense k-mesh confirms that
both U=0.5 and 1.5 eV calculations show the band gaps of 0.01 and 0.08 eV
respectively.

\begin{figure}[tbp]
\includegraphics [height=2.5in,width=3.2in] {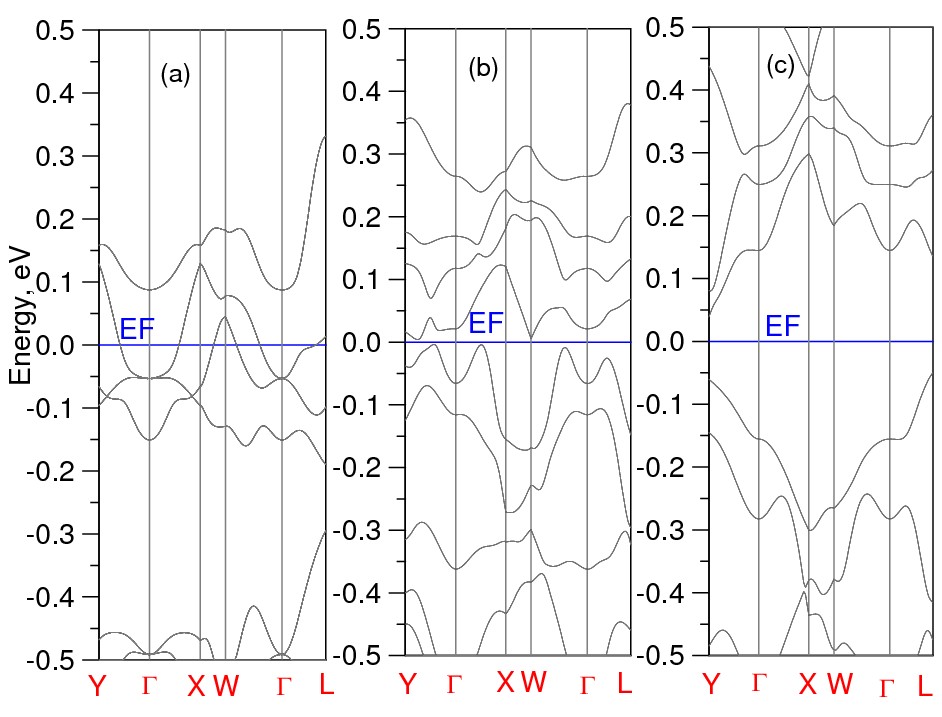}
\caption{Electronic band structure of CaOs$_{2}$O$_{4}$ shown along high
symmetry direction. (a) LDA+SO; (b) LSDA+SO+U, with U=0.5 eV; (c) LSDA+SO+U,
with U=1.5 eV}
\end{figure}

In order to examine whether the system becomes an Axion insulator, we recall
that similar to the Fu and Kane criterion\cite{FuKane}, which is designed
for non--magnetic systems with inversion symmetry, one can still use the
parity eigenvalues for the states at eight Time Reversal Invariant Momenta
(TRIM) in order to classify the magnetic insulators with inversion symmetry
and evaluate the magneto--electric coupling\cite{AriZhangAV,Bernevig}. 
\begin{equation}
\theta=\pi\cdot M(\func{mod}\ 2)  \label{1}
\end{equation}
where $M=(\sum_{k}N_{i})/2$, and $N_{i}$ is the number of occupied states at
the TRIM points $i$ with odd parity. This allows us to analyze their
topological electronic properties based on the parities of occupied bands.
One must also verify the Chern number of the bands vanish, to define the
magneto-electric coupling, which can be done by studying the evolution of
the band structure from a trivial insulating state. The local environment of
Os ion in spinels is oxygen octahedron, and the crystal--field splitting of
Os 5\textit{d} states is large, which makes the e$_{g}$ band\ higher (about
2 eV) than the Fermi level. The bands close to the Fermi level are mainly
contributed by t$_{2g}$ mixed with O 2\textit{p }orbitals. There are 4 Os
per unit cell, thus the number of t$_{2g}$ bands is 24. If all of 24 t$_{2g}$
bands are fully occupied, the system cannot possess any topologically
non--trivial properties. Noticing that Os occurs in its 3$^{+}$ valence, for
an insulator to occur, there are 20 occupied and 4 empty t$_{2g}$ bands.
Thus, instead of analyzing 20 occupied bands, we alternatively look at the
parities of 4 empty bands.

For the spinel lattice, the eight TRIM points are $\Gamma$, $X,\ Y,$ $Z$,
and four $L$ points. For the $\Gamma$ point, all of the t$_{2g}$ bands
possess even parity, while for\ $X,\ Y,$ and $Z$ points, two of the 4 empty
bands have even parity and the other two bands possess odd parity. Thus,
regardless the value of \textit{U}, the summation of the number of empty t$%
_{2g}$ states with odd parities at $\Gamma,$\ $X,\ Y,$ and $Z$ points is 6.
On the other hand, the Coulomb interaction has a significant effect on the
parities of the bands around the L points. This is shown in Table III, where
the summation of the number of the odd states are 6 and 4 for U=1.5 and 0.5
eV, respectively. Therefore, according to Eq.\ref{1}, for U=1.5 eV, we have $%
\theta=0$, which corresponds to a normal insulator. However, for U=0.5 eV,
we have $\theta=\pi $, which leads us to the axion insulator with novel
magnetoelectric properties.

\begin{table}[ptb]
\caption{Calculated parities of states at Time Reversal Invariant Momenta
(TRIMs) of CaOs$_{2}$O$_{4}$. Only the 4 empty t$_{2g}$ bands are shown in
order of increasing energy. Lx 2$\protect\pi/a$(-0.5,0.5,0.5), Ly 2$\protect%
\pi/a$(0.5,-0.5,0.5), Lz 2$\protect\pi/a$(-0.5,0.5,0.5) and L 2$\protect\pi%
/a $(0.5,0.5,0.5)).}%
\begin{tabular}{lllll}
\hline
& Lx & Ly & Lz & L \\ \hline
U=0.5 eV & ++++ & ++++ & ++++ & - - - - \\ 
U=1.5 eV & -+++ & -+++ & -+++ & + - - - \\ \hline
\end{tabular}%
\end{table}

Since topological insulators must be separated from trivial insulators by a
semi--metallic state\cite{Dirac-Metal}, with 3D Dirac like dispersion, we
additionally perform calculations for a number of intermediate values of 
\textit{U} to find out the boundary between semi-metal and Axion insulator
as well as between semi-metal and Mott insulator. Our calculations for 
\textit{U's} varying from 1.0 to 1.3 eV show that there are 3D Dirac
crossings close to (0.02, x, 0.02)2$\pi/a$ points of the BZ. where the value
of x will change with U. With both decreasing and increasing the \textit{U},
the Dirac point moves and annihilates by meeting with other Dirac points,
thus opening the energy gap and forming either the $\theta=\pi$ Axion
insulator or normal insulator, respectively. To summarize, our electronic
phase diagram (see Fig. 3) for $U<$0.4 eV, predicts CaOs$_{2}$O$_{4}$ to be
a metal; for 0.4$<U<$0.9 eV, an axion insulator; for 0.9$<U<$1.4 eV, a
topological semi-metal metal; for $U>$1.4 eV, a Mott insulator. Note, unlike
in the cubic pyrochlore iridates \cite{Dirac-Metal}, where the topological
semi-metal has Dirac points exactly at the Fermi energy, here, because of
lower symmetry of the magnetic order, the Dirac points are slightly above or
below the Fermi energy. To see the sensitivity of these results to the
lattice parameters we perform calculations with increasing and decreasing
the volume by 6\%, as well as adjusting the internal coordinates and
performing small changes in Os-O-Os bond angle. We confirm that while there
are some little changes in calculated energy bands, our predictions on Axion
insulator behavior are robust.

Replacing Ca by Sr, both the Os-O-Os bond angle and the bond length will
change as shown in Table I. However, same as CaOs$_{2}$O$_{4}$, SrOs$_{2}$O$%
_{4}$ shows the same rich phase diagram as a function of \textit{U}.

We finally study BaOs$_{2}$O$_{4}$ which has both the largest bond angle and
bond length as shown in Table I. These differences significantly affect its
band structure: the LDA+SO\ calculation with U=0 gives that at the $\Gamma$
point the eight J$_{eff}=1/2$\ states have degeneracies 4-2-2 and not 2-4-2
as found in Ca and Sr cases. Note that this sequence of levels is the same
as recently suggested by Pesin and Balents to realize topological insulating
scenario in pyrochlore iridates \cite{Pesin-Balents}. Unfortunately, the
bands crossing the Fermi level exist and cannot be removed by slight
adjusting the lattice constant. Same with CaOs$_{2}$O$_{4}$ and SrOs$_{2}$O$%
_{4}$, considering the Coulomb interaction \textit{U} will induce magnetism,
but we do not find axion-insulator or Dirac-metal state for BaOs$_{2}$O$_{4}$
for any reasonable \textit{U}.

\begin{figure}[tbp]
\includegraphics [height=1.5in,width=3.0in] {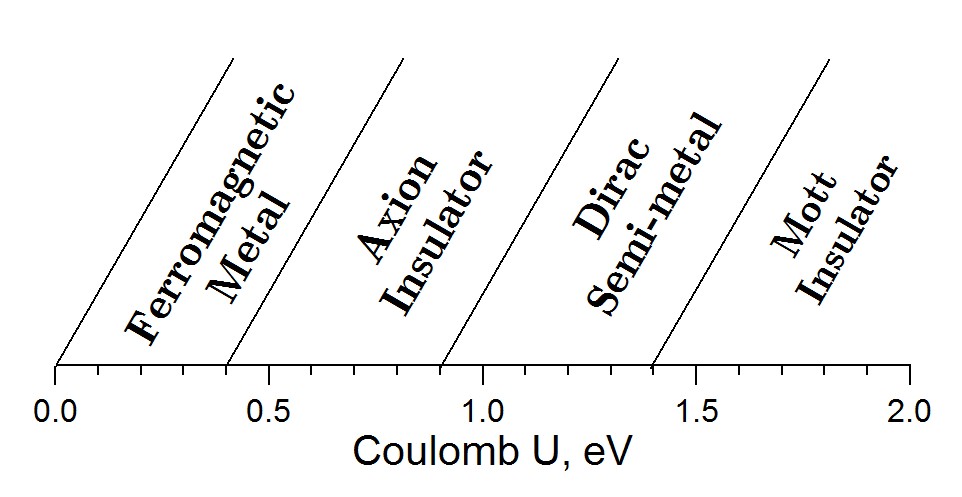}
\caption{Sketch of the predicted phase diagram for spinel osmates.}
\end{figure}

To summarize, using density functional based electronic structure
calculations, we have found that depending on the strength of the Coulomb
correlation among the 5d orbitals several exotic electronic phases may be
realized in Os based spinel compounds. In particular a magnetic topological
insulating phase (axion insulator) may be realized in CaOs$_{2}$O$_{4}$ and
SrOs$_{2}$O$_{4}$ with large orbital magneto--electric parameter $\theta=\pi$%
. This research suggests that new functionalities such as controlling
electrical\cite{TeoKane} and optical \cite{Axion-1} properties via magnetic
textures, can be found in the new 5d spinel materials that we propose.

X.W. acknowledges support by National Key Project for Basic Research of
China (Grant No. 2011CB922101, and 2010CB923404), NSFC under Grant No.
10974082. S.S. acknowledges support by DOE SciDAC Grant No.
SE-FC02-06ER25793 and thanks Nanjing University for the kind hospitality
during his visit to China. The work at Berkeley was supported by the Office
of BES, Materials Sciences Division of the U.S. DOE under contract No.
DE-AC02-05CH1123.

\end{document}